\documentclass[%
 reprint,
superscriptaddress,
 amsmath,amssymb,
 aps,
 prl,
]{revtex4-2}

\usepackage{graphicx}
\usepackage{color}
\usepackage{dcolumn}
\usepackage{bm}

\begin{document}

\preprint{APS/123-QED}

\title{Blue phosphorene bilayer is a two-dimensional metal - and an unambiguous classification scheme for buckled hexagonal bilayers}

\author{Jessica Arcudia}
\affiliation{Departamento de Física Aplicada, Centro de Investigación y de Estudios Avanzados, Unidad Mérida, Mérida, Yucatán, México}%

\author{Roman Kempt}
\affiliation{Technische Universität Dresden, Fakultät für Chemie und Lebensmittelchemie, Bergstraße 66c, 01062 Dresden, Germany}

\author{Miguel Eduardo Cifuentes-Quintal}%
\affiliation{Departamento de Física Aplicada, Centro de Investigación y de Estudios Avanzados, Unidad Mérida, Mérida, Yucatán, México}%

\author{Thomas Heine}
\email{thomas.heine@tu-dresden.de}
\affiliation{Technische Universität Dresden, Fakultät für Chemie und Lebensmittelchemie, Bergstraße 66c, 01062 Dresden, Germany}
\affiliation{Helmholtz Zentrum Dresden-Rossendorf, Leipzig Research Branch, Permoserstr. 15, 04318 Leipzig, Germany}
\affiliation{Department of Chemistry, Yonsei University, Seodaemun-gu, Seoul 120-749, Republic of Korea}

\author{Gabriel Merino}%
 \email{gmerino@cinvestav.mx}
\affiliation{Departamento de Física Aplicada, Centro de Investigación y de Estudios Avanzados, Unidad Mérida, Mérida, Yucatán, México}


\date{\today}

\begin{abstract}
High-level first-principles computations predict blue phosphorene bilayer to be a two-dimensional metal. This structure has not been considered before and was identified by employing a block-diagram scheme that yields the complete set of five high-symmetry stacking configurations of buckled honeycomb layers, and allows their unambiguous classification. We show that all of these stacking configurations are stable or at least metastable configurations both for blue phosphorene and gray arsenene bilayers. 
For blue phosphorene, the most stable stacking configuration has not yet been reported, and surprisingly it is metallic, while all other arrangements are indirect band gap semiconductors.
As it is impossible to interchange the stacking configurations by translations, all of them should be experimentally accessible via the transfer of monolayers. 
The metallic character of blue phosphorene bilayer is caused by its short interlayer distance of 3.01 {\AA} and offers the exceptional possibility to design single elemental all-phosphorus transistors.
\end{abstract}

\maketitle


\section{\label{sec:level1}Introduction}
Since the first exfoliation of graphene and the discovery of its remarkable properties \cite{novoselov2004electric}, many other 2D materials came in the focus of interest, including its isoelectronic hexagonal congener boron nitride \cite{britnell2012electron}, and the wide variety of transition metal dichalcogenides \cite{miro2014atlas,an2018exfoliation,haastrup2018computational}. 
However, it was not before the exfoliation black phosphorus \cite{liu2014phosphorene,li2014black}, that the attention for 2D pnictogens emerged. Black phosphorus is the most stable allotrope of phosphorus and crystallizes in an orthorhombic structure. But a single layer of phosphorus, so-called phosphorene, can also crystallize in other forms. The honeycomb structure predicted by Zhu and Tománek \cite{zhu2014semiconducting} is just 2 meV/atom higher in energy than the black counterpart, and given that its band gap value is slightly above the photon energy of visible blue light, it was named blue phosphorene (in the following bluP). This prediction was materialized just a couple of years later by Zhang et al., who successfully synthesized monolayer (ML) bluP by epitaxial growth on an Au(111) substrate \cite{zhang2016epitaxial}. The next pnictogen of interest is arsenic, mostly found in its bulk form as gray arsenic, and whose layered rhombohedral structure makes it an excellent candidate for exfoliation. The first studies on the stability and the properties of a single arsenic layer, i.e., gray arsenene (in the following grAs), predicted by Kamal and Ezawa \cite{kamal2015arsenene} and Kou et al. \cite{kou2015structural}, encouraged experimental groups to exfoliate few-layers arsenic \cite{beladi2019atomically}.

Both bluP and its arsenic congener grAs have a structure similar to graphene, which is a hexagonal lattice with their atoms alternatingly being displaced out of the 2D plane (Figure \ref{fig1}a). These monolayers show exciting electronic properties with an indirect band gap in the range of 1.5-2.0 eV \cite{zhu2014semiconducting,zhang2016epitaxial,kamal2015arsenene,beladi2019atomically}. BluP is a $p$-type semiconductor with high carrier mobility \cite{xiao2015theoretical}. On the other hand, grAs could be used for transistors or mechanical sensors, due to its indirect-direct band gap transitions \cite{kou2015structural,zhang2015atomically}, semiconductor-metal transitions \cite{zhu2014semiconducting,kamal2015arsenene}, and topological phase transitions under strain \cite{zhang2015quantum,wang2016tunable}.

It is well-known that interlayer interactions can significantly alter the properties of 2D materials, most strikingly discussed recently for superconductivity in bilayer (BL) graphene \cite{cao2018unconventional,cao2018correlated}, but also for band gap nature and valleytronics in transition-metal dichalcogenides (for a review, see Ref. \cite{kuc2015electronic}), or for metal-insulator transitions in noble metal chalcogenides (for a review, see Ref. \cite{kempt2020two}). The properties crucially depend on the stacking type and twist angle, which can be controlled using various transfer techniques \cite{novoselov20162d}.

Only a few studies addressed interlayer effects in bluP and grAs, and no details on the stacking order have been reported experimentally up to now \cite{zhang2016epitaxial,beladi2019atomically}. 
Given the corrugation of the monolayer, more than the high-symmetry AA and AB stacking orders are expected. In how many ways is it possible to stack them? Even for the simplest case of two buckled honeycomb layers, this question has not been answered yet. Herein, we suggest a new approach based on a simple graphical analysis, which we will call a \lq block diagram\rq\, to ascertain it.

Using block diagrams, we identify all possible high-symmetry stacking configurations of BL buckled honeycomb lattices, which are expected to yield all low-energy forms for bluP and grAs bilayers. 

Employing first principles calculations (density-functional theory (DFT), the Random Phase Approximation (RPA), and single-particle Green's function approach $G_0W_0$), we calculated their structures, thermodynamic and kinetic stabilities, and explore their electronic structures. We discovered that the lowest-energy bluP BL was not yet reported to date and, surprisingly, it is metallic.

\section{Results and Discussion}
Because both bluP and grAs crystallize in the trigonal lattice, we can describe a single layer in the subperiodic layer group $\mathbf{P}$3 (\#65) with the unique sites $1a = (0,0,z)$, $1b = (1/3, 2/3, z_{1})$ and $1c=(2/3, 1/3, z_{2})$, where two sites are occupied with $z_{1}=-z_{2}$ \cite{aroyo2006}. 

\begin{figure}
\includegraphics{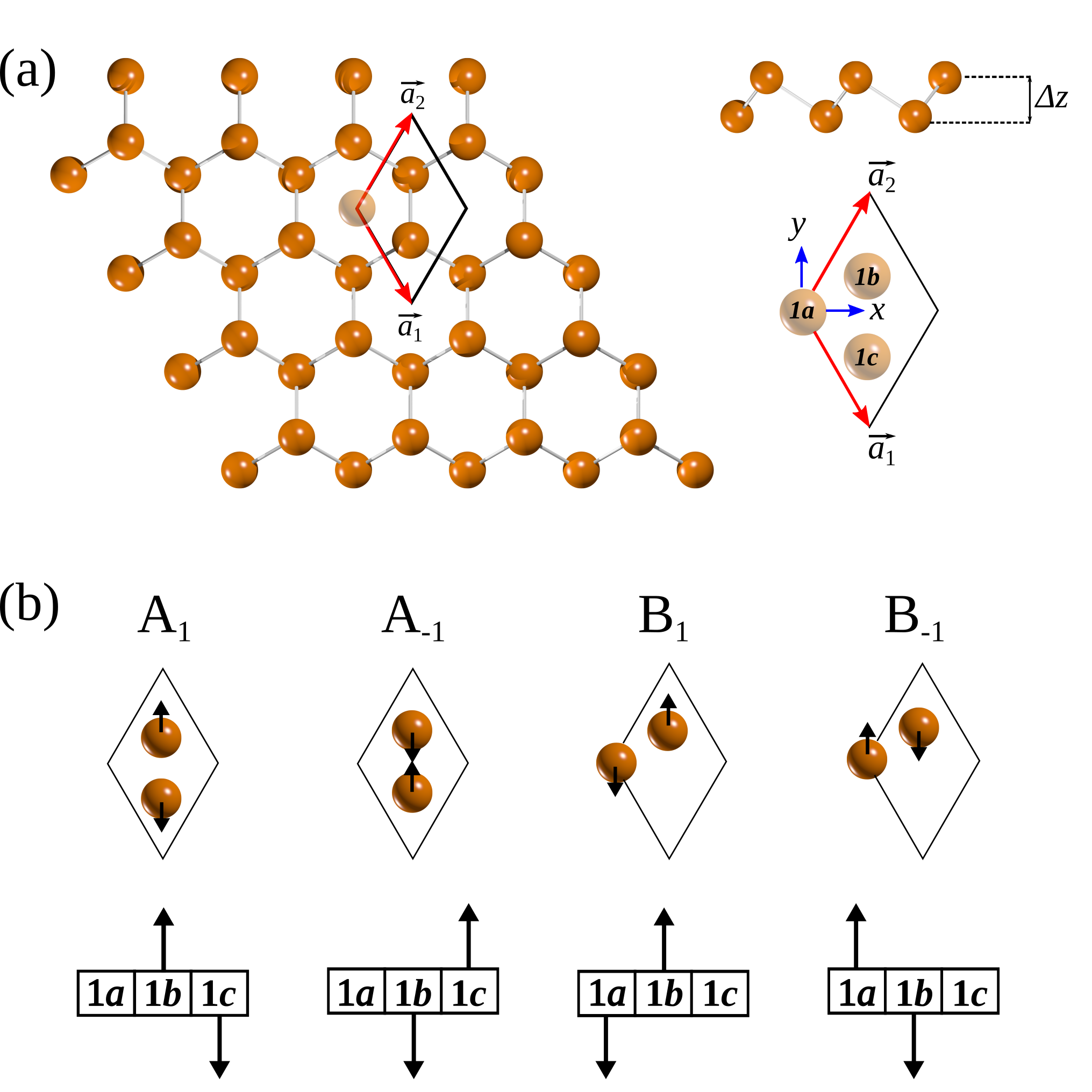}%
\caption{\label{fig1}(a) Front and side view of a buckled honeycomb monolayer. The unit cell with its lattice vectors and high symmetry sites 1a, 1b, and 1c. (b) Top: Four different representations of a unit cell (arrows indicate the out of plane displacement of the atoms), bottom: block diagrams to represent each monolayer.}
\end{figure}

The lattice vectors ($\vec{a}_{1}$ and $\vec{a}_{2}$) delimiting the two-dimensional lattice are defined as $\vec{a}_{1}=\frac{1}{2}a\hat{x}-\frac{\sqrt{3}}{2}a\hat{y}$, $\vec{a_{2}}=\frac{1}{2}a\hat{x}+\frac{\sqrt{3}}{2}a\hat{y}$, where $a$ is the lattice constant. Then the unit cells of any trigonal monolayer with a buckling can be represented as in Figure \ref{fig1}b, 
where the arrows indicate the out-of-plane displacements. We conveniently label these forms as A$_{1}$, A$_{-1}$, B$_{1}$, and B$_{-1}$, where the negative sign in the subscripts denotes a displacement change of the atoms with respect to the plane (up or down). These four configurations are symmetry-equivalent and can be used for the construction of the BL forms. For a better understanding, we design a simple graphical method, which we call \lq block  diagram\rq\, to facilitate the visualization.

\begin{figure}
\includegraphics{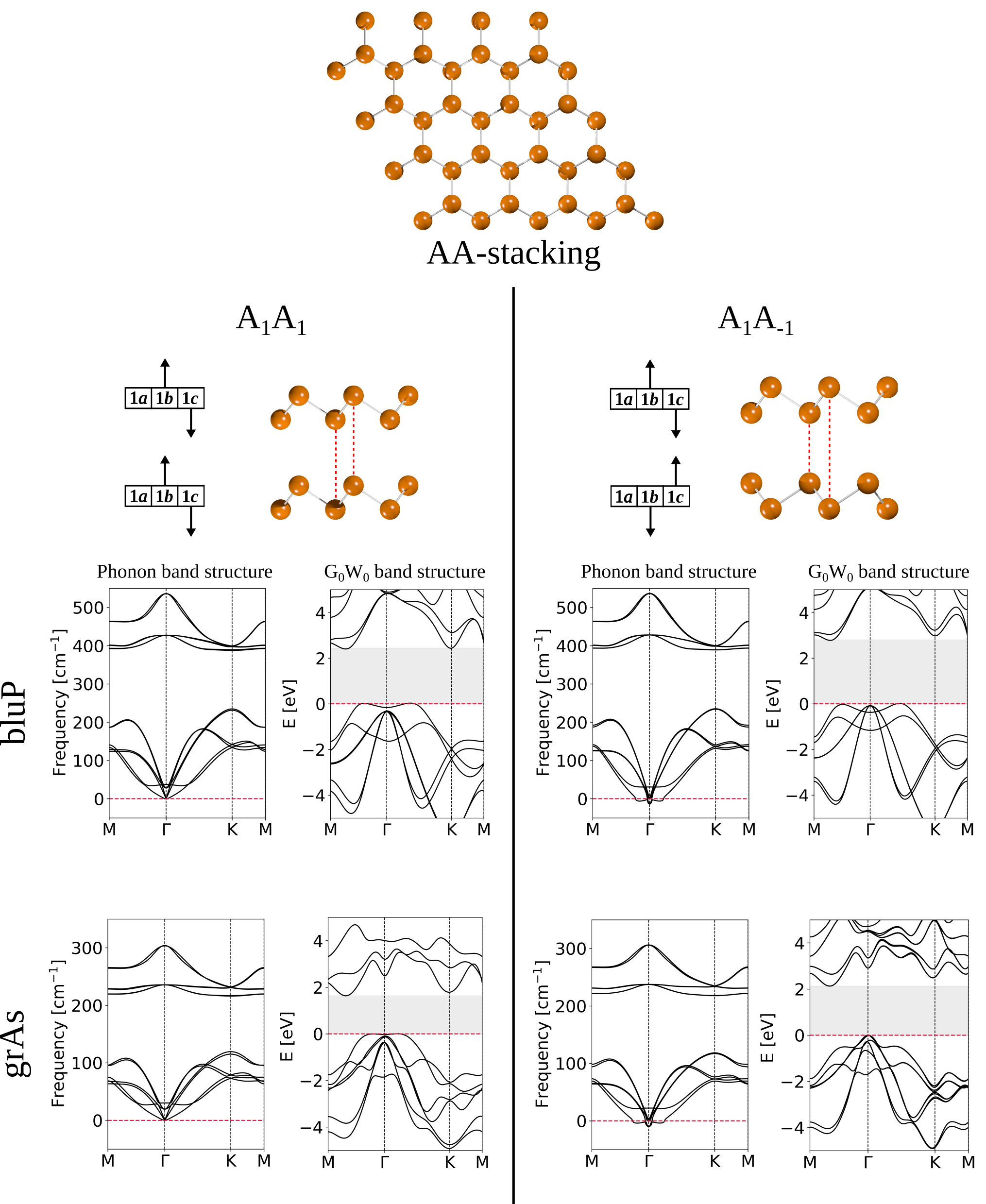}%
\caption{\label{fig2}Top view of an AA-type bilayer. The block diagram, atomistic structure, and phonon and electronic band structures of its two configurations: A$_{1}$A$_{1}$, and A$_{1}$A$_{-1}$}
\end{figure}

\begin{figure}
\includegraphics{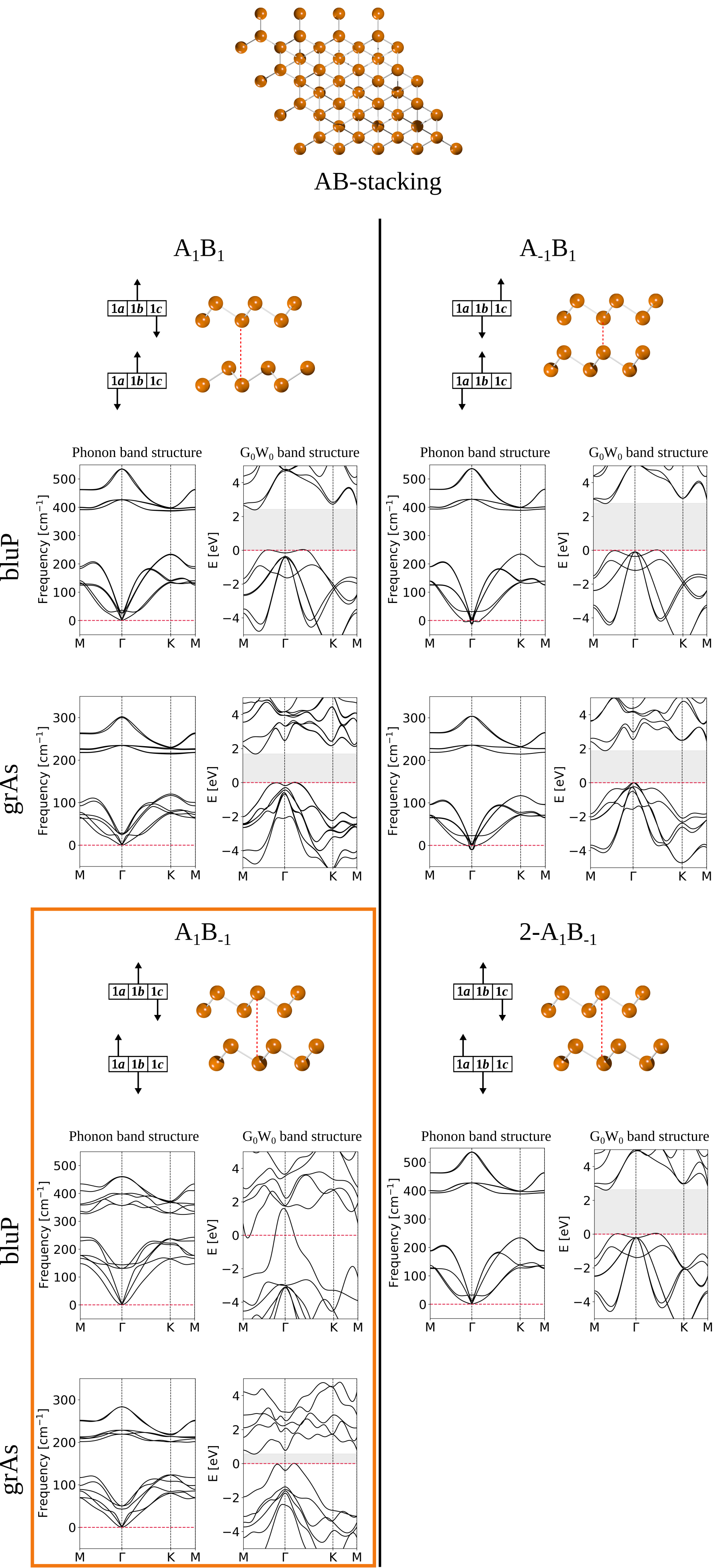}%
\caption{\label{fig3} Top view of an AB-type bilayer. The block diagram, atomistic structure, and phonon and electronic band structures of its three configurations: A$_{1}$B$_{1}$, A$_{-1}$B$_{1}$, and A$_{1}$B$_{-1}$. The second minimum of A$_{1}$B$_{-1}$ (2-A$_{1}$B$_{-1}$) is included. The most stable stacking configuration is indicated by a frame.}
\end{figure}

This block diagram is divided into three parts, and each of them specifies a site of the unit cell (1a, 1b, or 1c). The atoms occupy two of the three sites and the direction of the arrows is related to the buckling. So, to build a bilayer, one puts another block on the top. In this way, we identify the total number of different high-symmetry forms of corrugated honeycomb bilayer as five, where the stacking sequence of two of them is of AA-type, and for the remaining three it is AB-type (see block diagrams in Figure \ref{fig2} and \ref{fig3}). These five stacking configurations are reported for silicene, whose layers are covalently bound \cite{padilha2015} .

For BL bluP, only four different stacking configurations have been studied to date \cite{pontes2018layer,ahn2018phase}. First principles calculations indicated the A$_1$A$_1$ configuration to be the most stable one, followed by A$_1$B$_1$, A$_{-1}$B$_1$, and A$_1$A$_{-1}$. For grAs, there is an on-going debate whether the A$_1$B$_{-1}$ or the A$_1$A$_1$ form is the most stable one \cite{cao2015electronic,ahn2018phase,kecik2016stability,kadioglu2018diffusion}.

For all stable and metastable stacking configurations discussed in this work, full geometry optimizations yield similar structural parameters, independent on the density-functional, the choice of London dispersion correction scheme, the orbital representation (local basis functions vs. plane waves), or the underlying code (see Methods for details). In all but one case a single local minimum per stacking configuration is found. Only for the A$_{1}$B$_{-1}$ configuration of bluP we found two local minima: one minimum corresponds to a structure where the interlayer distance $d$ is small (3.01~{\AA}) and the structure shows small corrugation, corresponding to a smaller buckling height $\Delta z$. We call this structure A$_{1}$B$_{-1}$. The second structure (2-A$_{1}$B$_{-1}$) shows a larger interlayer distance of 4.93~{\AA}, corresponding to weakly interacting layers, and larger buckling. 
Both for bluP and grAs BL the closer interlayer distance in A$_{1}$B$_{-1}$ is accompanied by a lattice constant increase of $~$0.1~{\AA} and by a buckling reduction of $\Delta z \approx 0.07$ {\AA}.
The remainder of the structures has almost identical lattice constants and buckling heights as the monolayer ($a$=3.26~{\AA} and $\Delta z$=1.24 {\AA}, Table \ref{table1}) \cite{zhang2016epitaxial,wang2018electronic}.
Structural parameters for all stacking configurations of BL bluP and grAs are summarized in Table~\ref{table1}. 

\begin{table}
\caption{\label{table1}The lattice constant ($a$), buckling height ($\Delta z$), and interlayer distance ($d$), computed at the PBE+MBD level for blue phosphorene (bluP) and gray arsenene (grAs) bilayers.}
\begin{ruledtabular}
\begin{tabular}{c c c c c c c}
 & \multicolumn{3}{c}{bluP BL} & \multicolumn{3}{c}{grAs BL}\\
System & $a$ ({\AA}) & $\Delta z$ ({\AA}) & $d$ ({\AA}) & $a$ ({\AA}) & $\Delta z$ ({\AA}) & $d$ ({\AA})\\
\hline
A$_{1}$B$_{-1}$ & 3.36 & 1.17 & 3.01 & 3.69 & 1.35 & 3.66 \\
A$_{1}$A$_{1}$ & 3.26 & 1.24 & 4.66 & 3.60 & 1.40 & 4.58\\
A$_{1}$B$_{1}$ & 3.27 & 1.24 & 4.68 & 3.61 & 1.40 & 4.38\\
2-A$_{1}$B$_{-1}$ & 3.27 & 1.24 & 4.93 & - & - & -\\
A$_{-1}$B$_{1}$ & 3.26 & 1.24 & 5.36 & 3.59 & 1.40 & 5.43\\
A$_{1}$A$_{-1}$ & 3.26 & 1.24 & 5.42 & 3.59 & 1.40 & 5.48\\
Monolayer & 3.26 & 1.24 & - & 3.60 & 1.40 & -\\
\end{tabular}
\end{ruledtabular}
\end{table}

All investigated BL systems are significantly more stable than their ML counterparts and are unlikely to exfoliate without severe intrusion. For both bluP and grAs, the A$_1$B$_{-1}$ stacking configuration was found to be the most stable one. In both cases, this structure has distinct features making it quite different compared to the other configurations: It has the smallest interlayer distance and the smallest corrugation. The interlayer binding energy, defined as $E_{ib}=\frac{E_{BL}-2E_{ML}}{N},$ exceeds 180 meV per atom in both cases (Table \ref{table2}). It is important to note that local and hybrid density-functionals give different stacking orders for the stacking configurations, which result in disagreement on the most stable form. Substantiation at the RPA level, independent if starting from a PBE or PBE0 calculation, result in the same stacking order and clearly identify A$_1$B$_{-1}$ to be the most stable form. 

For bluP A$_1$B$_{-1}$ BL, we show an abrupt change in buckling height $\Delta z$ and the distance between layers $d$ upon variation of lattice constant $a$, corresponding to the phase transition from the metallic configuration A$_{1}$B$_{-1}$ to the semiconducting configuration 2-A$_{1}$B$_{-1}$ (see Fig. S1a-b in the Supplementary Material). In Fig. S1c we plot the buckling height $\Delta z$ vs.~the energy difference $\Delta E$, indicating that  A$_{1}$B$_{-1}$ is energetically favorable as it is less corrugated. Fig. S1d indicates that A$_{1}$B$_{-1}$ is extremely sensitive to variations in the interlayer distance $d$. 2-A$_1$B$_{-1}$ shows weaker interlayer interactions and is located in the second, very shallow minimum.

While for BL bluP the configuration A$_{1}$B$_{-1}$ is by far the most stable one, for BL grAs two configurations are energetically competitive: A$_1$A$_1$ and A$_1$B$_1$ are only less than 26 meV per atom higher in energy. Relative energies with respect to the monolayers and with respect to the most stable bilayer forms are given in Table~\ref{table2}. 

\begin{table} 
\caption{\label{table2}The relative energies ($\Delta E$) and the interlayer binding energy ($E_{ib}$) for all hexagonal blu phosphorene and gray arsenene stacking configurations, computed at the RPA+rSE@PBE0 level (including the ZPE correction). Units are in meV/atom.}
\begin{ruledtabular}
\begin{tabular}{c c c c c}
 &  \multicolumn{2}{c}{bluP BL} & \multicolumn{2}{c}{grAs BL}\\
 System & $\Delta$E & E$_{ib}$ & $\Delta$E & E$_{ib}$\\
 \hline
A$_{1}$B$_{-1}$ & 0.0 & -180.2 & 0.0 & -185.2 \\
A$_{1}$A$_{1}$ & 83.9 & -96.2 & 25.7 & -159.5 \\
A$_{1}$B$_{1}$ & 86.8 & -93.4 & 16.1 & -169.0\\
2-A$_{1}$B$_{-1}$ & 104.1 & -76.1 & - & - \\
A$_{-1}$B$_{1}$ & 126.2 & -54.0 & 108.9 & -76.3\\
A$_{1}$A$_{-1}$ & 128.4 & -51.8 & 96.5 & -88.6 \\
\end{tabular}
\end{ruledtabular}
\end{table}

The phonon dispersion of the bluP and grAs MLs show the typical features of 2D materials, with two linear and one out-of-plane quadratic branches of the acoustic modes, and a clear energetic separation of optical and acoustic branches (Figure S2). For bluP and grAs BL systems (Figure \ref{fig2} and \ref{fig3}), the phonon dispersion significantly differs between the most stable A$_{1}$B$_{-1}$ and the other structures: While for all systems the three acoustic modes are in the same energy range as three low-energy optical modes (emerging from the 2nd bilayer system and indicating weaker interactions), for A$_{1}$B$_{-1}$ the stronger interlayer interaction lifts the low-energy optical modes towards higher energies and these branches show significantly less dispersion compared to the other stacking configurations. These shifted vibrational modes are of E$_g$ and A$_g$ symmetry and Raman active, and thus could serve for the characterization of this BL, similar as it has been demonstrated for graphene and silicene experimentally \cite{liang2017low}. For the high-energy arrangements A$_{-1}$B$_{1}$ and A$_{1}$A$_{-1}$ of both bluP and grAs BLs we found small imaginary frequencies which are due to the limitations of the numerical approach used in the phonon calculations.

To substantiate the electronic band gaps, the electronic structures have been recalculated within the quasiparticle approach, using the single-shot $G_{0}W_{0}$ approximation on top of PBE Kohn-Sham bands. Spin-orbit coupling (SOC) was included in all calculations for the grAs systems (see Table~\ref{table3}). Besides the well-known band gap underestimation, the limitations of local DFT calculations also include the incorrect location of the valence band maximum (VBM) for bluP (see also Figure S3) \cite{iyikanat2019}.

Most importantly, both DFT and $G_{0}W_{0}$ identify bluP BL in the most stable configuration A$_{1}$B$_{-1}$ as a metal. All other systems are indirect band gap semiconductors, with the $G_{0}W_{0}$ band gaps being considerably larger compared to those calculated by PBE. For BL bluP, electronic band gaps range from 2.39 to 2.82 eV, values that are lower than those of the ML (3.25 eV). For grAs, the A$_{1}$B$_{-1}$ low-energy form has a remarkably small band gap of 0.58 eV, while the other systems range from 1.38 eV to 2.04 eV, thus being somewhat narrower than that of the ML (2.30 eV). 
For all semiconducting bluP and grAs BLs we found an interesting competition of a parabolic band climaxing at the $\Gamma$ point and a Mexican hat structure around $\Gamma$ point to become the VBM. The latter is obtained for bluP BL in A$_1$A$_1$ and A$_1$B$_1$ configuration (somewhat less pronounced for grAs BL). A similar Mexican hat structure has been reported for GaSe and other III-VI monolayers, and the small dispersion may be useful for small FET structures and could give rise to Landau levels \cite{aziza2018valence,kuc2017high}.

\begin{table} 
\caption{\label{table3}Indirect band gaps (E$_{gap}$) of ML and BL stacking configurations of bluP and grAs calculated at PBE and $G_{0}W_{0}$ levels of theory. Units are in eV. Spin-orbit coupling (SOC) was considered for the grAs systems.}
\begin{ruledtabular}
\begin{tabular}{c c c c c}
   &  \multicolumn{2}{c}{E$_{gap}$-bluP} & \multicolumn{2}{c}{E$_{gap}$-grAs}\\
System & PBE & $G_{0}W_{0}$ & PBE-SOC & $G_{0}W_{0}$-SOC\\
\hline
A$_{1}$B$_{-1}$ & - & - & 0.12 & 0.58 \\
A$_{1}$A$_{1}$ & 1.23 & 2.39 & 0.84 & 1.64 \\
A$_{1}$B$_{1}$ & 1.25 & 2.40 & 0.63 & 1.38 \\
2-A$_{1}$B$_{-1}$ & 1.41 & 2.59 & - & - \\
A$_{-1}$B$_{1}$ & 1.58 & 2.80 & 1.27 & 2.04 \\
A$_{1}$A$_{-1}$ & 1.60 & 2.82 & 1.22 & 1.99 \\
Monolayer & 1.96 & 3.25 & 1.43 & 2.30 \\
\end{tabular}
\end{ruledtabular}
\end{table}

Finally, we return to the remarkable 2D metal A$_{1}$B$_{-1}$ bluP. The band structure (Figure \ref{fig3}) includes metallic bands crossing the Fermi level near the $\Gamma$ point, characterizing it as conventional metal. In addition, there is a basin of charge carriers close to $M$ point. This interesting band structure suggests anisotropic electronic properties that shall not be further explored at this stage. 

\section{Summary and Outlook}
We present a scheme to identify and to label all symmetrically distinct stacking configurations of corrugated honeycomb bilayers and investigate the resulting structures blue phosphorene (bluP) and gray arsenene (grAs). We discovered a new, and a the same time the most stable, configuration of bluP, which has a small interlayer distance, quenched corrugation and is a new member of the exclusive group of two-dimensional metals. 

Besides the most stable A$_1$B$_{-1}$ form, also A$_1$A$_1$ and A$_1$B$_1$ stacking configurations could be obtained experimentally by layer transfer techniques, as their large interlayer interaction and symmetry constraints prevent interconversion or relaxation into other structures. Lower-symmetry configurations as they are known for BL graphene are unlikely here due to the surface corrugation. This holds both for bluP and grAs. Except for the metallic bluP A$_1$B$_{-1}$ form, all investigated bluP and grAs bilayers are indirect band gap semiconductors and resemble the Mexican hat-type feature of the electronic bands near the valence band maximum which have been intensively discussed for III-VI-group mono- and multilayer systems.

BluP shows a semiconductor-metal transition when going from the ML to the most stable BL. Such transitions have been found for other 2D materials (e.g. noble metal chalcogenides \cite{miro2014two} and GeP$_3$ \cite{jing2017gep3}) and could be used to design single-material transistors with low Schottky barrier between the electrode and semiconducting scattering region \cite{ghorbani2016single}. In a similar vein, we think it is a remarkable challenge to construct the first single-element transistor based on blue phosphorus, with its metallic bilayer configuration serving as electrode material.

\section{Methods}
To identify all potentially existing isomeric structures, we started geometry optimization with different starting structures, varying interlayer distances and buckling.
All geometries have been fully optimized by means of DFT within the framework of the projector-augmented wave (PAW) method \cite{blochl1994projector,kresse1999ultrasoft} as implemented in the Vienna ab initio simulation package (VASP 5.4.4) \cite{kresse1996efficiency,kresse1996efficient}. 

The valence states were expanded in plane-waves with an energy cutoff of 400 eV. The Perdew, Burke, and Ernzerhof (PBE) exchange-correlation functional was employed \cite{perdew1996generalized}, with London dispersion interactions taken into account by the many-body dispersion (MBD) correction as suggested by Tkatchenko and coworkers \cite{tkatchenko2012accurate}. For comparison, we also used the Barone and Adamo's hybrid PBE0 functional \cite{perdew1996rationale,adamo1998toward} which does not affect the geometries of the bilayers.
The first Brillouin zone was sampled using a $\Gamma$-centered k-grid of 15x15x1 and 42x42x1 k-points for hexagonal phosphorene and arsenene bilayers, respectively, and Gaussian smearing of 0.10 eV. These numerical parameters were chosen to ensure a convergence criterion of 10$^{-8}$ eV in total energy, and 10$^{-3}$ eV/{\AA} in atomic forces. A vacuum space of 25{\AA} was considered in order to avoid interactions with the repeated images. The lattice parameters of each system were obtained by direct minimization of the total energy, with the atomic positions fully optimized until the interatomic forces were less than 10$^{-3}$ eV/{\AA}. 

The same level was employed for the phonon calculations, for which we used the small displacement method as implemented in the Phon code \cite{phon}. The force constant matrix was computed using central differences within atomic displacements of 0.02{\AA} in 9x9 supercells.

Band structures have been calculated with $G_{0}W_{0}$ \citep{hedin1965,shishkin2006kresse}, considering spin-orbit coupling (SOC) for all arsenic systems.

Relative stabilities have been calculated from single-point energies of optimized structures on the grounds of PBE-MBD and PBE0-MBD, which give different stacking orders and required calculations beyond DFT (see Table S1). We employed the Random-Phase Approximation (RPA) with the renormalized single excitation correction (rSE) \cite{ren2013renormalized} based on Kohn-Sham orbitals from the PBE and PBE0 level as implemented in FHI-AIMS \cite{blum2009ab} on tight tier 1 numeric atom-centered orbitals with added auxiliary diffuse basis functions on 12x12x1 $k$-grids to determine the most stable form. The RPA stacking orders are independent on the choice of the underlying density-functional (for comparison, see Supplementary Material), herein we discuss only the RPA+rSE$@$PBE0 variant. Relative energies $\Delta E$ (in meV per atom) are given with respect to the most stable stacking configuration, and interlayer binding energies $E_{ib}$ are defined with respect to the monolayers.

\section{Acknowledgements}
This work was funded in Mexico by Cinvestav (Grant SEP-Cinvestav-2018-57). The CGSTIC (Xiuhcoatl) is acknowledged for allocation of computational resources. J. A. thanks Conacyt for her Ph.D. fellowship.
We thank the Center for Information Services and High Performance Computing (ZIH) at TU
Dresden for generous allocation of computer time. We thank BMBF (NobleNEMS) for financial support.
We thank Andras Kis (EPFL) for fruitful discussions on 2D heterostructure manufacturing techniques.

\bibliographystyle{apsrev4-2}
\bibliography{References}

\end{document}